\documentclass[aps,floats,twocolumn,prb,showpacs]{revtex4-1}
\usepackage{graphicx}
\usepackage{color}
\usepackage{upgreek}
\usepackage{setspace}
\usepackage{amssymb}
\usepackage{amsmath}
\usepackage{pstricks}
\usepackage{dsfont}
\usepackage{fancyhdr}
\usepackage{array}
\usepackage{multirow}
\usepackage{placeins}
\begin{document}
\title{Nonlocal correlations and spectral properties of the Falicov-Kimball model}
\author{T. Ribic$^1$, G. Rohringer$^{1,2}$, and K. Held$^1$}
\affiliation{$^1$Institute of Solid State Physics, TU Wien, 1040
Vienna, Austria \\
$^2$Russian Quantum Center, Novaya street, 100, Skolkovo, Moscow region 143025, Russia}
\date{Version 0.95, preview version \today}

\begin{abstract}
We derive an analytical expression for the local two-particle vertex of the Falicov-Kimball model, including its dependence on all three frequencies, the full vertex and all reducible vertices. This allows us to calculate the self energy in diagrammatic extensions of dynamical mean field theory, specifically in the dual fermion and the one-particle irreducible approach.  Non-local correlations are thence included  and originate here from charge density wave fluctuations. At low temperatures and in two dimensions, they lead to a larger  self energy contribution at low frequencies and a more insulating spectrum.
\end{abstract}

\pacs{71.27.+a, 71.10.Fd}
\maketitle

\let\n=\nu \let\o =\omega \let\s=\sigma  
\newcommand{\Gred}{\overline{G}}
\newcommand{\Chired}{\overset{\sim}{\chi}}

% 71.27.+a  Strongly correlated electron systems; heavy fermions
% 71.10.Fd  Lattice fermion models (Hubbard model, etc.)
% 71.30.+h  Metal-insulator transitions and other electronic transitions
% 71.20.Eh  Rare earth metals and alloys
% 75.20.Hr  Local moment in compounds and alloys; Kondo effect, valence
%           fluctuations, heavy fermions
% 75.47.Gk  Colossal magnetoresistance

\section{Introduction}
\label{Sec:Intro}
%%%%%%%% begin kh
In 1969 the Falicov-Kimball model (FKM)  \cite{Falicov69} 
was introduced for describing 
 SmB$_6$ and its semiconductor-to-metal transition.
Falicov and Kimball considered fully immobile, Sm-$f$ electrons, interacting with mobile conduction electrons. Nowadays we know that the  FKM does not describe the Kondo physics that is so important for metallic $f$-electron systems since it requires at least a minimal  $f$-electron mobility or spin-flip. 
Since Plischke \cite{Plischke72} also showed that the  paramagnetic metal-insulator transition in the coherent potential approximation (CPA) is  a smooth crossover rather than a phase transition, interest in the FKM faded in the 1970s.

Interest resurfaced in the 1980s when it was realized \cite{Kennedy86} that the FKM is a  simplified version of the   Hubbard model \cite{Hubbard63}  and arguably the simplest model for electronic correlations.  This often allows for analytical solutions. An important analytical result was achieved in 1986 when Brandt and Schmidt \cite{Brandt86} and, independently, Kennedy and  Lieb \cite{Kennedy86} proved that there is a phase transition towards a checkerboard  charge density wave (CDW) of the mobile and, antithetically, immobile electrons for dimension $d\geq 2$.  Freericks and coworkers showed rigorously that alongside the CDW there is 
phase separation in the limit of small \cite{Freericks96} and large interaction strength \cite{Freericks02}.

The dawn of dynamical mean field theory (DMFT) \cite{Metzner89a,MuellerHartmann89,Georges92a,Jarrell92a} saw a further rapid development  for the FKM.
Among others, Brandt and Mielsch \cite{Brandt89} solved the paramagnetic FKM exactly within DMFT or for dimension $d\rightarrow \infty$;
van Dongen and Vollhardt \cite{vanDongen90} studied CDW order; Freericks and Miller \cite{Freericks00} determined dynamical and transport properties; and Jani\v{s} proved the equivalence to the CPA solution of the FKM.\cite{Janis91} For a concise review of the DMFT results we refer the reader to  Ref.\  \onlinecite{Freericks03} by Freericks and Zlati\'c.

The FKM remains an interesting physical model for mixed valence systems and binary alloys, and an ideal testbed for analytical results and new approaches. Regarding the latter, we have seen  considerable efforts to include non-local correlations beyond DMFT. These started with the $1/d$ approach
\cite{Schiller95}  and cluster extensions of DMFT \cite{DCA,clusterDMFT,LichtensteinDCA,Potthoff,Maier04} and have been applied to the FKM by Schiller\cite{Schiller99} and Hettler {\em et al.} \cite{Hettler00}, respectively.

More recently, diagrammatic extensions of DMFT became the focus of this methodological development. These extensions start with a local two-particle vertex  \cite{vertex} and from this construct the local DMFT correlations as well as non-local correlations beyond.
Different flavours of these diagrammatic approaches are the
  dynamical vertex approximation 
(D$\rm \Gamma$A) \cite{DGA1,Katanin09}, cf. Ref.\  \onlinecite{Kusunose06,DGAintro}, the  dual fermion (DF) approach  \cite{DualFermion} and non-local expansion \cite{Li15}, the one-particle irreducible approach (1PI), \cite{1PI} the merger of DMFT with the functional renormalization group (DMF$^2$RG)  \cite{DMF2RG},  the  triply-irreducible local expansion \cite{Ayral15} and DMFT+fluctuation exchange (FLEX) \cite{Kitatani15}. Diagrammatic extensions of the CPA on the basis of the parquet approach  have been introduced in Ref.\ \onlinecite{Janis01b}. Among others, these approaches allowed to calculate the critical exponents in the Hubbard model \cite{Rohringer11,Hirsch15} and FKM  \cite{Antipov14}. In agreement with the expectation from universality, these exponents are of Heisenberg- and Ising-type, respectively.

In the present paper,  we derive an analytical expression for the full vertex of the mobile electrons, including its Matsubara frequency $\omega = 0$ component, for the irreducible vertices in the particle-hole and particle-particle channel as well as for the fully irreducible vertex, employing the parquet equation. The vertices irreducible in given channels have been known before, see e.g.\ Ref.\ \onlinecite{Freericks03,Janis14,Shvaika}. 

These local vertices of the DMFT solution are the starting point of the aforementioned diagrammatic extensions of DMFT, and hence an analytical expression is most valuable. Here, we employ the ladder series in the particle-hole channel to derive 
 analytical expressions for the DF and 1PI self-energy. We present explicit results for the paramagnetic self-energy and spectral function when approaching the CDW transition of the two dimensional Falicov-Kimball model and discuss the differences between DF and 1PI.
Our results for the spectral evolution complement the pioneering work by Antipov {\em et al.}  \cite{Antipov14} which focused instead on the DF critical exponents for the CDW phase transition. Let us also mention the seminal work by Jani\v{s} and Pokorn\'y\cite{Janis10} and Pokorn\'y and Jani\v{s}\cite{Pokorny13} studying vertex corrections to the conductivity.
\\In Section \ref{Sec:LocalVertexAd} we present the expressions for the local, DMFT vertices: the full vertex, the irreducible vertices in the particle-hole and particle-particle channel and the fully irreducible one. This is supplemented in Section \ref{numResDMFT} by numerical results for these vertices at two different interaction strengths. From these local DMFT vertices we calculate in Section \ref{AnDerCorTer} the DF and 1PI self-energy which includes non-local correlations beyond DMFT. Section \ref{NumRes2D} shows numerical results obtained this way for the two-dimensional Falicov-Kimball model.
Finally, Section \ref{Conclusio} summarizes our main findings. 
%% end kh
\section{Local vertex functions for the Falicov-Kimball model}
\label{Sec:LocalVertex}
\subsection{Analytic derivation of local vertex functions}
\label{Sec:LocalVertexAd}
%%beg tr
The Hamiltonian of the one-band spin-less Falicov-Kimball model reads
\begin{equation}
 \label{equ:deffk}
 \hat{\mathcal{H}}=-t\sum_{\langle ij \rangle}\hat{c}^{\dagger}_i\hat{c}^{\phantom{\dagger}}_j+U\sum_i\hat{c}^{\dagger}_i\hat{c}^{\phantom{\dagger}}_i\hat{f}^{\dagger}_i\hat{f}^{\phantom{\dagger}}_i-\mu\sum_i\hat{c}^{\dagger}_i\hat{c}^{\phantom{\dagger}}_i-\varepsilon_f\sum_i\hat{f}^{\dagger}_i\hat{f}^{\phantom{\dagger}}_i,
\end{equation}
where $\hat{c}_{i\sigma }^{\dagger}$($\hat{c}_{i\sigma }$) creates
(annihilates) an {\sl itinerant} electron at lattice site $i$ and $\hat{f}^{\dagger}_i$($\hat{f}_i$) creates (annihilates) a {\sl localized} electron at lattice site $i$; $t$ denotes the hopping amplitude of itinerant electrons between nearest-neighbours, and $U$ is the local Coulomb interaction between an itinerant and a localized electron on the same lattice site $i$; $\mu$ and $\varepsilon_f$ are the local
potentials for the itinerant and localized electrons respectively, subsuming the chemical potential. In the following
$\beta\!=\!1/T$ denotes the inverse temperature. For the case of a two-dimensional square-lattice considered for the numerical results we choose $D=4t\equiv 1$ as unit of energy. Our analytical equations are, with an appropriate lattice summation and dispersion relation $\varepsilon_k$, valid for any FKM, but of course the calculated local vertex is within the DMFT approximation and the non-local correlations beyond DMFT rely on the DF or 1PI approximations.

Let us recall that for the Falicov Kimball model the DMFT solution for the one-particle Green's function of the itinerant electrons can be found (semi-)analytically since the solution of the corresponding impurity model [i.e., the resonant level model (RLM)] can be obtained explicitly:\cite{Freericks03}
\begin{equation}
 \label{equ:rlm}
 G^{(1)}(\nu)\equiv G_{loc}(\nu)=p_1\underset{{\cal G}^U(\nu)}{\underbrace{\frac{1}{{\cal G}^{-1}(\nu)-U}}}+p_2{\cal G}(\nu),
\end{equation}
where $p_1=\langle\hat{f}^{\dagger}_i\hat{f}^{\phantom{\dagger}}_i \rangle$, $p_2=1-p_1$, and ${\cal G}(\nu)$ is the local, non-interacting Green function of the RLM. 
From Eq. (\ref{equ:rlm}) the DMFT self-energy $\Sigma(\nu)= {\cal G}^{-1}(\nu)-G(\nu)^{-1}_{loc}(\nu)$ can be obtained and the lattice Dyson equation   $ G_{loc}(\nu)=\sum_{\mathbf{k}} G_{DMFT}(\nu,\mathbf{k})$ closes the DMFT self-consistency cycle. Here $\varepsilon_{\mathbf{k}}=-2t(\cos k_x+\cos k_y)$ is the dispersion of the square lattice, $\sum_{\mathbf{k}}\equiv 1/(2\pi)^2\int_{-\pi}^{\pi}dk_xdk_y\;$ and $G_{DMFT}(\nu,\mathbf{k})=1/[i\nu+\mu-\varepsilon_{\mathbf{k}}-\Sigma(\nu)]$. Specifically, $\Sigma(\nu)$ reads in terms of ${\cal G}(\nu)$ [or ${\cal G}^U(\nu)$]:
\begin{equation}
 \label{equ:sigmarlm}
 \Sigma(\nu)=\frac{p_1U}{1-p_2U{\cal G}(\nu)}
\end{equation}

Similar as for the DMFT self-energy, in the Falicov-Kimball model the local DMFT vertex functions for the itinerant electrons can be calculated (semi-)analytically. The reason is the non-interacting nature of this system: The localized electrons can be seen as just an additional potential for the itinerant ones [see Eq. (\ref{equ:rlm})] which are otherwise non-interacting. The impurity solution in Eq.\ (\ref{equ:rlm}) is simply the sum of two terms, with and without a present $f$-electron (potential).
As will be shown in the following this leads to a factorization of the vertex functions in terms of one-particle quantities. (self-energy $\Sigma(\nu)$) 

\begin{figure}
\includegraphics[width=.9\linewidth]{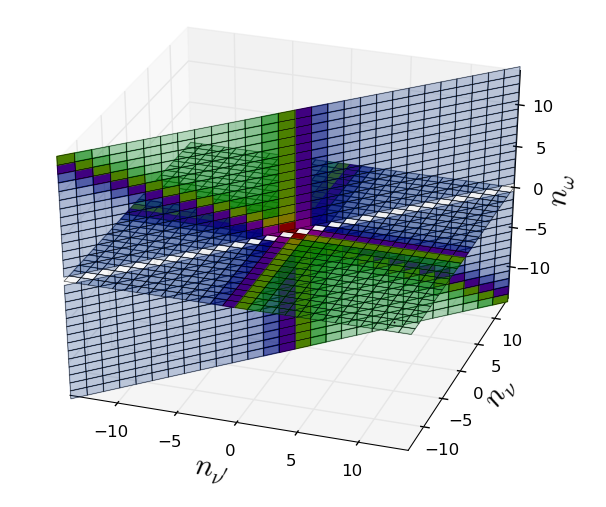}
\caption{\label{SchemaF}Schematic representation of the structure of the full vertex $F$  in (Matsubara) frequency space [fermionic Matsubara frequencies: $\nu_{n_\nu}=(2n_{\nu}+1)\pi/\beta$; bosonic ones $\omega_{n_{\omega}}=2n_{\omega}\pi/\beta$].   There are only $\omega=0$ and $\nu=\nu'$ contributions.}
\end{figure}

The vertex functions of the FKM only have a finite values for $\o = 0$ and for $\nu = \nu ' $ (note once again the similarity with a non-interacting system). This corresponds to the fact, that $c$-electrons are unable to exchange energy directly between themselves. They can only scatter {\sl indirectly} via the $f$ electrons. However, no energy transfer can occur  in such processes due to the zero-bandwidth of the $f$-electrons. The reduced structure is depicted schematically in figure \ref{SchemaF}.
The simple structure as well as the factorization property of the one-particle irreducible (1PI) vertex $F$ allows for an explicit calculation of the irreducible vertices in the particle-particle ($\Gamma_{pp}$), particle-hole ($\Gamma_{ph}$) and transverse particle-hole ($\Gamma_{\overline{ph}}$) and by extension, the fully irreducible vertex $\Lambda$ (for the definitions of $F$, $\Gamma_r$ and $\Lambda$ see e.g.\ Refs. \onlinecite{DGA1}, \onlinecite{vertex}). 

Let us start with the definition of the DMFT local two-particle Green's function of the FKM $G^{(2)}(\nu,\nu',\omega)$:
\begin{align}
\label{equ:defgreen2p}
G^{(2)} (\nu,\nu',\omega) = \int d\tau_1d\tau_2d&\tau_3\;e^{-i\nu\tau_1}e^{i(\nu+\omega)\tau_2}e^{-i(\nu'+\omega)\tau_3}\nonumber\\ &\times \left\langle\text{T}\left(\hat{c}^{\dagger}(\tau_1)\hat{c}(\tau_2)\hat{c}^{\dagger}(\tau_3)\hat{c}\right)\right\rangle,
\end{align}
where T denotes the time-ordering operator and $\langle\ldots\rangle$ the thermal average at the (inverse) temperature $\beta$.
The frequency convention here is chosen in accordance with particle-hole ($ph$) notation\cite{vertex}. 

As already mentioned, for the Falicov-Kimball model, the DMFT impurity problem actually consists of the weighted average of two non-interacting problems: one where no localized $f$-electron is present [${\cal G}(\nu)$ term in Eq. (\ref{equ:rlm})] and one where the existence of such localized electron generates a potential $U$ for the itinerant electrons [${\cal G}^U(\nu)$ term in Eq. (\ref{equ:rlm})]. For each of these two non-interacting situations Wick's theorem holds and allows us to express the two- (and multi-)particle Green's functions in terms of the one-particle ones:
\begin{align}
\label{equ:vertrlm}
G^{(2)} ( \nu , &\nu' , \omega ) = \beta (\delta_{\omega , 0} - \delta_{ \nu , \nu' })\times\nonumber\\& \left[p_1G^{U}_{0} (\nu) G^{U}_{0} (\nu'+\omega) +p_2G_{0} (\nu) G_{0} (\nu' + \omega)\right],
\end{align}
with ${\cal G}^{(U)}(\nu)$ being defined in Eq. (\ref{equ:rlm}).

\begin{figure}
\includegraphics[width=.9\linewidth]{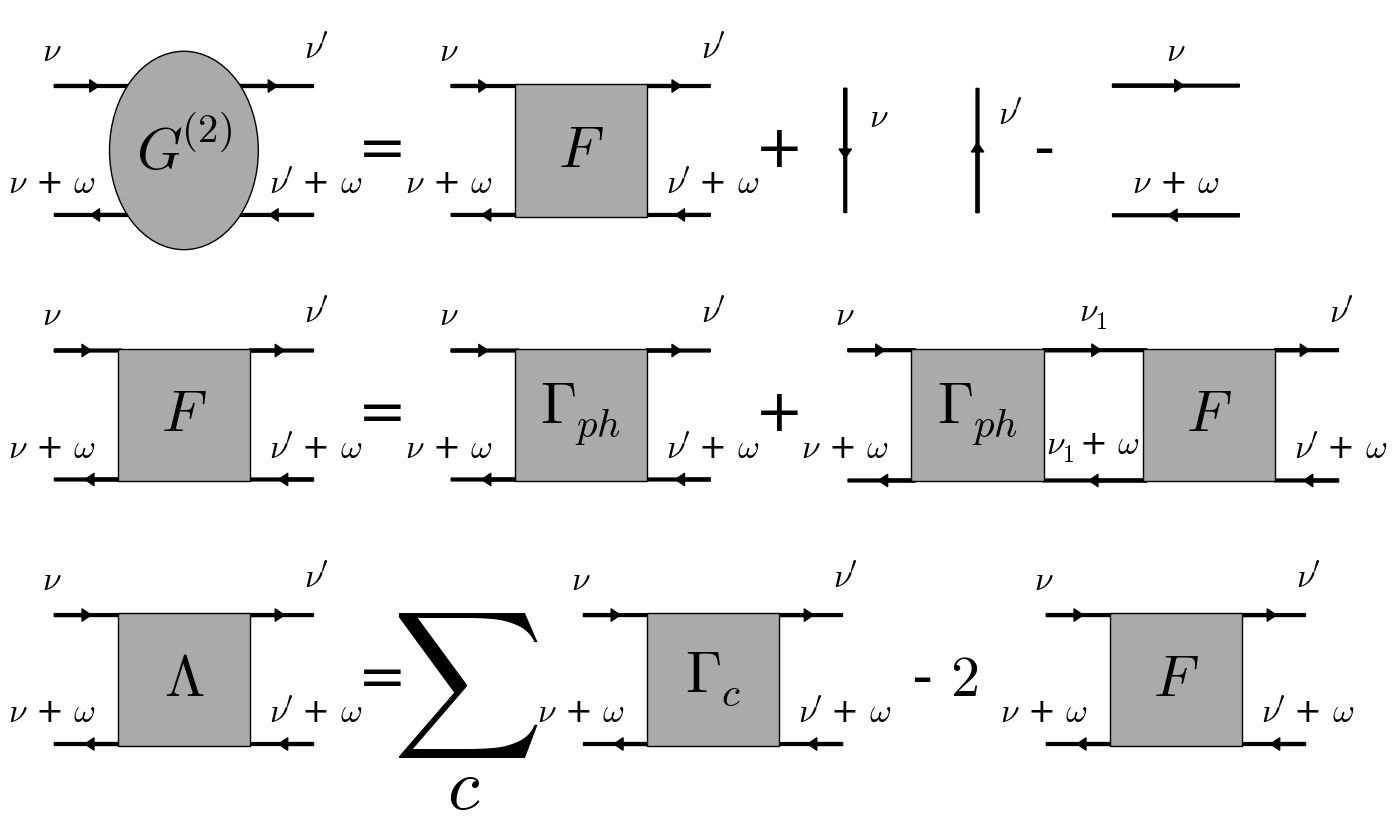}
\caption{\label{Feyndiags}Feynman-diagrammatic relation between the two-particle Green's function $G^{(2)}$ and the full vertex $F$ (top), $F$ and the irreducible vertex in the $ph$ channel $\Gamma_{ph}$, given by the Bethe-Salpeter equation (middle), and the relation between $F$, $\Gamma_c$ and the fully irreducible vertex $\Lambda$ (bottom). }
\end{figure}

The two-particle Green's function can be decomposed into a disconnected and a connected part containing the full (one-particle irreducible) vertex function\cite{vertex} $F^{\nu\nu'\omega}$, see fig. \ref{Feyndiags} (top) [note that $F$ is defined with opposite sign compared to Ref. \onlinecite{vertex}]:
\begin{multline}
\label{equ:deff}
G^{(2)}( \nu , \nu' , \omega ) = \beta (\delta_{\omega , 0} - \delta_{ \nu , \nu' }) G(\nu) G(\nu' + \omega) + \\ G(\nu) G(\nu+\omega) F^{ \nu  \nu' \omega } G(\nu') G(\nu' + \omega) .
\end{multline}
From Eqs. (\ref{equ:vertrlm}) and (\ref{equ:deff}) it is well illustrated that the two-particle local DMFT vertex of the FKM can be expressed exclusively in terms of the one-particle non-interacting Green's functions ${\cal G}(\nu)$ and ${\cal G}^U(\nu)$ or, equivalently, through the local DMFT self-energy $\Sigma(\nu)$. 
Using algebraic identities [including Eq. (\ref{equ:sigmarlm})] it is possible to express $F$ as:
\begin{equation}
\label{equ:frlm}
F^{\nu , \nu' , \omega} = \beta (\delta_{\omega , 0} - \delta_{ \nu , \nu' }) a(\nu) a(\nu' + \omega)
\end{equation} 
with 
\begin{equation}
\label{equ:defa}
a(\nu) = \dfrac{(\Sigma(\nu) - U) \Sigma (\nu) }{\sqrt{p_1 p_2} U}
\end{equation} 
This result has been already obtained previously\cite{Freericks03,Janis14}. 

The full vertex can be further decomposed into irreducible and reducible contributions in the $pp$, $ph$ and $\overline{ph}$ channels\cite{vertex}. The Bethe-Salpeter equations in the respective channels relate the full vertex $F$ and the vertices irreducible in the given channel $\Gamma_c$ ($c=ph,\overline{ph},pp$), see fig. \ref{Feyndiags} (middle):
\begin{multline}
\label{equ:bsph}
F^{\nu , \nu' , \omega} = {\Gamma_{ph}}^{\nu , \nu' , \omega} - \\ \dfrac{1}{\beta}  \sum_{\nu_1} F^{\nu , \nu_1 , \omega} G(\nu_1 + \omega) G (\nu_1) {\Gamma_{ph}}^{\nu_1 , \nu' , \omega}
\end{multline}
for the $ph$-irreducible vertex $\Gamma_{ph}$ and 
\begin{multline}
\label{equ:bspp}
F^{\nu , \nu' , \omega} = {\Gamma_{pp}}^{\nu , \nu' , \omega} + \dfrac{1}{2 \beta} \, \sum_{\omega_1} F^{\nu , \nu' + \omega - \omega_1, \omega_1 } \\ G(\nu + \omega_1) G (\nu' + \omega - \omega_1) {\Gamma_{pp}}^{\nu + \omega_1, \nu', \omega - \omega_1}
\end{multline}
for the $pp$-irreducible vertex $\Gamma_{pp}$ ($\Gamma_{\overline{ph}}$ can be obtained from $\Gamma_{ph}$ by means of the crossing symmetry\cite{vertex}). Regarding $F$'s special structure, these equations can be solved analytically for  $\Gamma_i$. We will start with the $ph$-vertex $\Gamma_{ph}$. First we consider the case $\omega \neq 0$. In this situation Eq. (\ref{equ:bsph}) is easily solved yielding
\begin{equation}
\label{equ:gammaph}
\Gamma_{ph}^{\nu , \nu' , \omega} = -  \delta_{\nu , \nu'} \dfrac{\beta a(\nu) a(\nu'+\omega) }{1 + G^{(1)}(\nu) G^{(1)} (\nu'+\omega) a(\nu) a(\nu'+\omega) }.
\end{equation}
 Let us stress that the expression obtained for $\Gamma_{ph}^{\nu\nu'(\omega \neq 0)}$ is consistent with the corresponding result obtained in \cite{Freericks03}. 

Calculating $\Gamma_{ph}^{\nu\nu'\omega}$ for $\omega = 0$ requires slightly more work. The resulting expression is less well known than the $\omega \neq 0$ one, but it has already been obtained by Shvaika\cite{Shvaika}. For the sake of brevity, we will consider the case where $\nu \neq \nu'$, but the obtained solution holds for all cases. Inserting the expression for $F$ [Eq. (\ref{equ:frlm})] in the Bethe-Salpeter equation (\ref{equ:bsph}) for $\omega=0$ yields:
\begin{multline}
\beta a(\nu) a(\nu') = \\ \Gamma_{ph}^{\nu , \nu', 0} - a(\nu) \sum_{\nu_1}  a ( \nu_1 ) G^{(1)} (\nu_1) G^{(1)} (\nu_1) \Gamma_{ph}^{\nu_1 , \nu', 0} \\  + a ( \nu ) a ( \nu ) G^{(1)} (\nu) G^{(1)} (\nu) \Gamma_{ph}^{\nu , \nu', 0}
\end{multline}
which can be rearranged to yield
\begin{multline}
\Gamma_{ph}^{\nu , \nu', 0} = \\ \beta \dfrac{a(\nu) \left( a(\nu') + 1/ \beta  \, \sum_{\nu_1} G^{(1)} (\nu_1) G^{(1)} (\nu_1) \Gamma_{ph}^{\nu_1 , \nu', 0}\right) }{ 1 + \, a ( \nu ) a ( \nu ) G^{(1)} (\nu) G^{(1)} (\nu) }
\label{factor}
\end{multline}
The right hand side of equation \ref{factor} factorizes into a part dependent on and a part independent of $\nu$ and, hence, the same holds for the left hand side, i.e. $\Gamma_{ph}^{\nu , \nu' , \omega}$. Thus making the ansatz $\Gamma_{ph}^{\nu\nu'(\omega=0)}=\beta C b(\nu)b(\nu')$ one can easily show that $b(\nu)$ is given by
\begin{equation}
b(\nu) =  \dfrac{a(\nu) }{ 1 +  a ( \nu ) a ( \nu ) G^{(1)} (\nu) G^{(1)} (\nu) },
\end{equation}
and the proportionality factor $C$ reads as
\begin{equation}
\label{equ:defC}
C = \left( 1 - \sum_{\nu_1} \dfrac{ \left( a ( \nu_1 )  G^{(1)} (\nu_1) \right)^2 }{ 1 + \left( a ( \nu_1 )  G^{(1)} (\nu_1) \right)^2}  \right)^{-1}.
\end{equation}
Summing up the results for $\omega\ne0$ and $\omega=0$ the vertex $\Gamma_{ph}^{\nu\nu'\omega}$ has the form:
\begin{multline}
\Gamma_{ph}^{\nu , \nu' , \omega} = \beta \delta_{\omega , 0} \, C \, b(\nu ) b(\nu' ) \\ - \delta_{\nu , \nu'} \dfrac{a(\nu)a(\nu'+\omega)}{1 + G^{(1)}(\nu ) G^{(1)} (\nu'+\omega)a(\nu')a(\nu'+\omega)} .
\end{multline}
Let us stress that the $\omega=0$ part of the irreducible vertex is of high relevance for the calculation of static susceptibilities in the framework of DMFT. 
It is also worth recalling that, for $\omega=0$, the occurrence of several\cite{VertDiv2} divergences in the irreducible vertex functions of DMFT have been reported both for the Falicov Kimball\cite{Janis01,Janis01b,VertDiv,Janis14,VertDiv2} and for the Hubbard model\cite{VertDiv,VertDiv2}. In fact, we also observe the occurrence of divergencies in $\Gamma_{ph}$: One can easily see that the prefactor $C$, defined in Eq. (\ref{equ:defC}), diverges once the sum over $\nu_1$ in this expression becomes equal to $1$. This leads to the divergence of the irreducible vertex function $\Gamma_{ph}^{\nu , \nu' , 0}$ at all fermionic frequencies $\nu,\nu'$. Fig.\ \ref{Vertdiv} shows a false color plot of $C$, displaying the positions of these vertex divergences in the DMFT phase diagram of the Falicov Kimball model.
%Let us also point out an interesting feature regarding the factor $C$ defined in Eq. (\ref{equ:defC}): One can easily see that the $C$-factor diverges once the sum over $\nu_1$ in this expression becomes equal to $1$ leading to a divergence of the irreducible vertex function. Such divergences have been observed in the Falicov-Kimball as well as the Hubbard model before\cite{Janis01,Janis01b,VertDiv,Janis14}. 
% ??? include in final version ??? Fig.\ 
%Fig. \ref{Vertdiv} shows a false color plot of  $C^2 / T$ throughout the phase diagram, indicating several lines of divergence of  $\beta C$ and hence the irreducible vertex function similar as found before for the Hubbard model in Ref.\ \onlinecite{VertDiv2}.

Let us now turn our attention to the vertex $\Gamma_{pp}$. It can be obtained from the full vertex $F$ by means of the Bethe-Salpeter equation for the particle-particle channel, Eq.\ (\ref{equ:bspp}). Using the explicit expression for $F$ in Eq. (\ref{equ:frlm})  Eq.\ (\ref{equ:bspp}) can be written as
\begin{multline}
\label{equ:bspp1}
\beta a(\nu) a(\nu' + \omega) ( \delta_{\omega , 0} - \delta_{ \nu , \nu'} ) =  \Gamma_{pp}^{\nu , \nu', 0} + \dfrac{1}{2} a(\nu) a(\nu' + \omega) \\ G^{(1)} (\nu) G^{(1)} (\nu' + \omega) \left( \Gamma_{pp}^{\nu , \nu' , \omega} - \Gamma_{pp}^{\nu' + \omega , \nu' , \nu - \nu'} \right)
\end{multline}
Using the crossing symmetry $F^{\nu , \nu' , \omega} = - F^{\nu , \nu + \omega , \nu' - \nu}$,
which also applies to the $pp$-irreducible vertex, one obtains for $\Gamma_{pp}$
\begin{multline}
\Gamma_{pp}^{\nu , \nu' , \omega} = \beta ( \delta_{\omega , 0} - \delta_{ \nu , \nu'} ) \\ \dfrac{a(\nu) a(\nu'')}{1 + a(\nu) a(\nu'') G^{(1)} (\nu) G^{(1)} (\nu'')}
\end{multline}

As $\Gamma_{\overline{ph}}$ can be deduced from $\Gamma_{ph}$ via the crossing symmetry  
\begin{equation}
\Gamma_{\overline{ph}}^{\nu , \nu' , \omega} = - \Gamma_{ph}^{\nu , \nu + \omega , \nu' - \nu},
\end{equation}
we are in a position to calculate the fully irreducible vertex $\Lambda$ via the Parquet equation\cite{vertex} 
\begin{equation}
\Lambda^{\nu , \nu' , \omega} = \sum_i \Gamma_{i}^{\nu , \nu' , \omega} - 2 F^{\nu , \nu' , \omega}
\end{equation}
with $i=ph,\overline{ph}$ or $pp$, respectively. Altogether we obtain
\begin{multline}
\label{equ:lambda}
\Lambda^{\nu , \nu' , \omega} = \beta a(\nu) a(\nu'') ( \delta_{\omega , 0} - \delta_{ \nu , \nu'} ) \\ \left( C \dfrac{b(\nu ) b(\nu'')}{a(\nu ) a(\nu'')} - 2 \dfrac{ a ( \nu )  G^{(1)} (\nu) a ( \nu'' )  G^{(1)} (\nu'') }{ 1 +  a ( \nu )  G^{(1)} (\nu) a ( \nu'' )  G^{(1)} (\nu'') } \right),
\end{multline}
completing the parquet decomposition of the local DMFT vertex of the FKM.
\vskip 10mm

\subsection{Local numerical results from DMFT}
\label{numResDMFT}

\begin{figure}
\includegraphics[width=.95\linewidth]{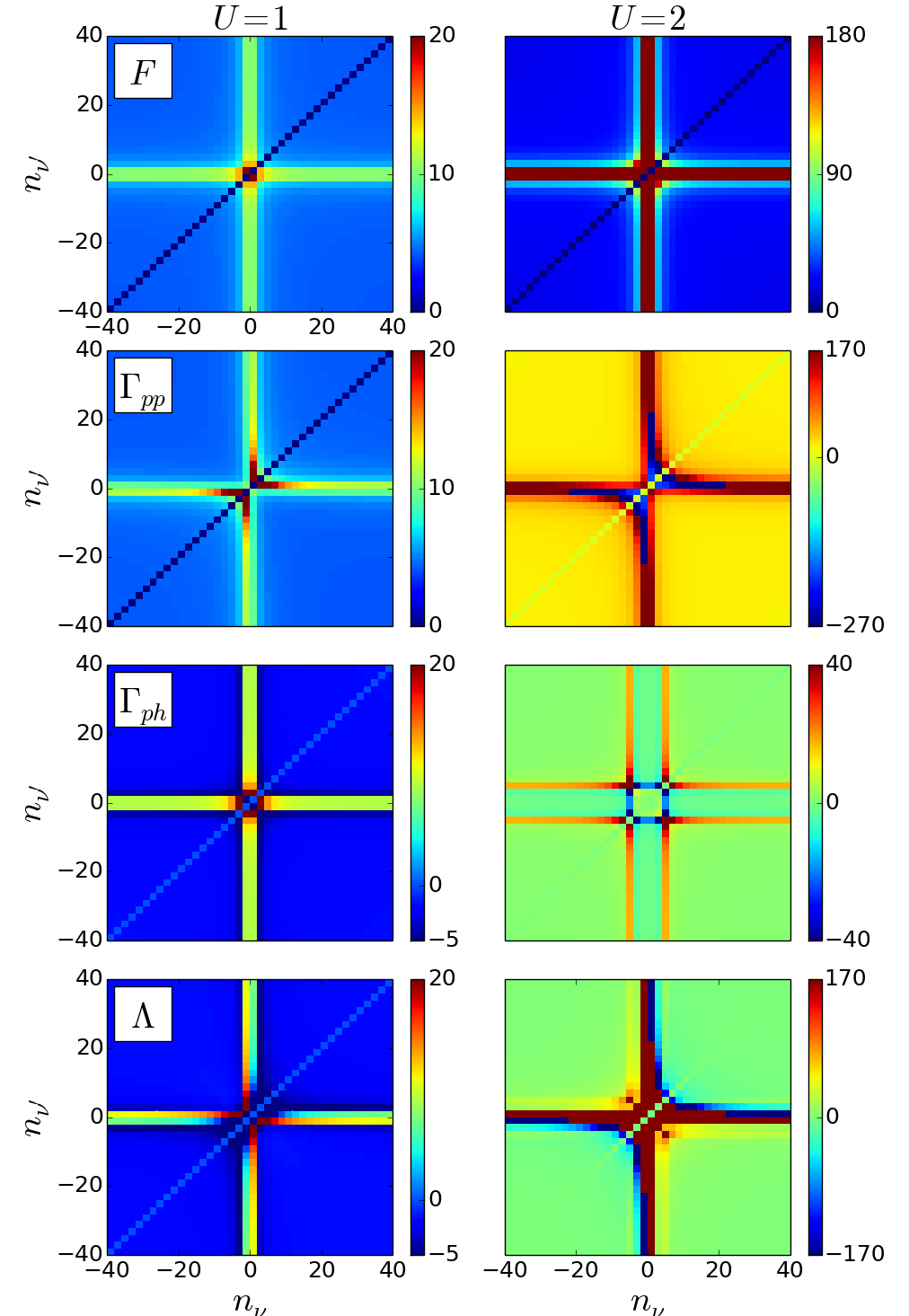}
%\caption{\label{FU1}Full vertex function for $\o = 0$ at $U=1$ and $T = 0.06?$}
\caption{\label{Allvert}
%{\bf ??? It would be better if the value 0 is set to white in these plots!} I can offer manually painting it white and mention that in the text
From top to bottom: full vertex, particle-particle-irreducible, particle-hole irreducible and fully irreducible vertex for $U=1$ (left) and $U=2$ (right) at $T = 0.06$ for $\o = 0$ for the Matsubara frequency indices of the incoming electron ($n_\nu$) and hole ($n_{\nu'}$) .}
\end{figure}

\begin{figure}
\includegraphics[width=.75\linewidth]{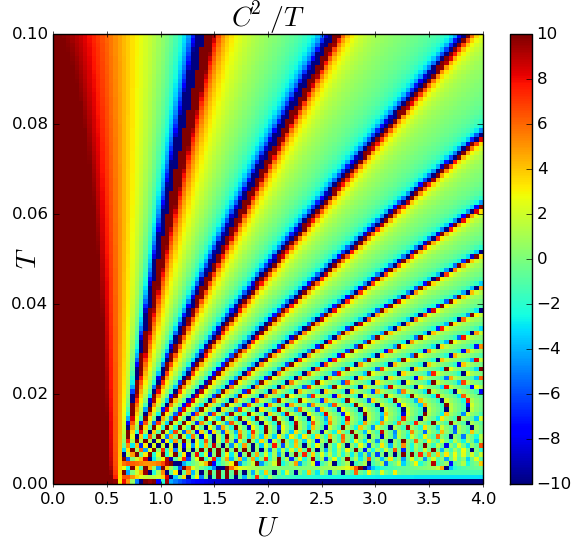}
\caption{\label{Vertdiv} Divergencies of the $C$-factor, Eq.\ \ref{equ:defC}, and hence the local, irreducible vertices  throughout the  $U$-$T$ phase diagram of the $2D$ Falicov-Kimball model. Note that the divergencies are visible in the picture as jumps from neagtive (blue) to positive (red) values. The red area on the left side for $U < 0.6$ is a consequence of the $1/T$ scaling and does not signify a divergency. In the lower right part, the lines are hardly visible due to emergent Moir\'e patterns.}
\end{figure}

In Fig. \ref{Allvert} the local DMFT vertex functions of the FKM are depicted for two values of the coupling, $U=1$ and $U=2$, respectively, for a half-filled lattice $p_1=p_2=0.5$ ($n_c=n_f=0.5$) at a fixed temperature $T=1/\beta=0.06$. Here, the vertex is depicted as density plot in the $\nu$-$\nu'$-plane for a fixed value of the bosonic Matsubara frequency. For our plots we choose $\omega=0$ which is arguably the most interesting case since for $\omega\ne0$ the vertex function consists only of another plane along $\nu=\nu'$. Let us also note that on a bipartite lattice at half-filling the vertex functions are purely real. 
%%% ???include in final version???
\begin{figure}
\includegraphics[width=.9\linewidth]{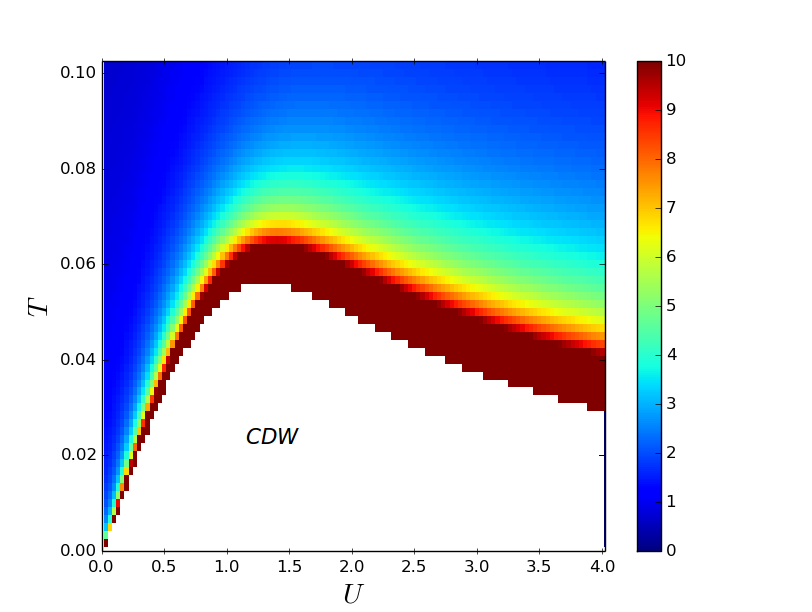}
\caption{\label{Phasediag}DMFT phase diagram for the $2D$ Falicov-Kimball model. Shown is the CDW susceptibility as a function of $U$ and $T$. In the region where $\chi(\omega=0,\mathbf{q}=(\pi,\pi))$ is negative -indicated by the white region in the phase diagram- the system is already in the CDW ordered phase.}
\end{figure}

The {\sl full vertex} is depicted in the first row of Fig. \ref{Allvert} for $U=1$ (first row, left panel) and $U=2$ (first row, right panel). In the half-filled case considered here the vertex $F^{\nu\nu'(\omega=0)}$ is a purely real function of its fermionic Matsubara frequencies $\nu$ and $\nu'$. The features which can be observed in the frequency structure of the vertex are the constant background, a diagonal where the amplitude of the vertex is $0$ and a cross-like structure in the center. Only  for very small values of the Matsubara frequencies one can see deviations from these main structures. The origin of the predominant features described above can be easily understood from Eqs. (\ref{equ:deff}) and (\ref{equ:defa}): At half-filling $\Sigma(\nu)=U/2+i\Sigma^{''}(\nu)$ where $\Sigma^{''}(\nu)$ denotes the imaginary part of the DMFT self-energy. Hence, the full vertex acquires the form:
\begin{align}
 \label{equ:fdiscuss}
  F^{\nu\nu'0}=&\left[\frac{U^2}{4}+\Sigma^{''}(\nu)^2+\Sigma^{''}(\nu')^2+\frac{4}{U^2}\Sigma^{''}(\nu)^2\Sigma^{''}(\nu')^2\right]\nonumber\\ &\times(1-\delta_{\nu,\nu'})\beta
\end{align}
{\sl First} of all, one can easily observe that the constant background is given by the term $U^2/4$ in Eq. (\ref{equ:fdiscuss}). This term can be interpreted as the bare part of the interaction between the itinerant electrons in the FKM which is mediated by the localized electrons and hence of the order $U^2$. Note that this is different for the corresponding vertex in the Hubbard model where the constant background of the vertex is just given by the interaction $U$ since the mobile particles have a direct interaction among themselves\cite{vertex}. {\sl Second} the zero values on the diagonal clearly arise from the factor $1-\delta_{\nu,\nu'}$ in the second line of Eq. (\ref{equ:fdiscuss}). This feature, which can be also observed in the triplet ($\uparrow\uparrow$) particle-particle irreducible vertex of the Hubbard model\cite{vertex}, can be interpreted as a consequence of the Pauli principle: for $\nu=\nu'$ both electrons would be in the same state which is forbidden by the Pauli principle. {\sl Third} the cross-like structure observed in the center extending to infinite values of the Matsubara frequencies originates from the second and the third terms ($\Sigma{''}(\nu)^2+\Sigma{''}(\nu')^2$) in Eq. (\ref{equ:fdiscuss}): E.g., for $\nu'=\pi/\beta$ the large contribution along this line in the frequency space stems from the rather large value $\Sigma^{''}(\pi/\beta)$ even for large values of $\nu$ where all the other contributions of the vertex (apart from the constant term) are suppressed at least as $1/\nu$. This explains the horizontal line of the cross-like structure. An analogous analysis for $\nu=\pi/\beta$ explains the vertical line. {\sl Finally} the last term of (the first line of) Eq. (\ref{equ:fdiscuss}) [$U^2/4\Sigma^{''}(\nu)\Sigma^{''}(\nu')$] decays in all directions of the frequency space and, hence, yields relevant contributions only within a small frequency box around the origin. 

In comparison with the corresponding full vertex of the Hubbard model\cite{vertex} one realizes that a large contribution at the secondary diagonal ($\nu=-\nu'$) is missing. This can be well understood from the fact that such features arise from the scattering events of two itinerant electrons on the same lattice site and at the same time. However, again such scattering events are not possible in the FKM due to the Pauli principle. 

Since the size of the main structures of the vertex functions scales with $\Sigma(\nu)$ it is clear that the vertex will become larger in parameter regimes where the self-energy is strongly enhanced. This is nicely illustrated by a comparison of the left ($U=1$) and right ($U=2$) panels in the first line of Fig. \ref{Allvert}: For the larger value of $U$ the corresponding self-energy is strongly enhanced since the system is in the insulating phase, where the self-energy is very large at low fermionic Matsubara frequencies. At $U=1$ on the other hand we are just at the verge of the metal-insulator transition.\cite{vanDongen97,Freericks03} The self-energy is still more moderate than for $U=2$, explaining the significantly smaller size of the vertex at this parameter.

A similar analysis as for the full vertex $F$ can be performed for the irreducible ones. As one can observe in Fig. \ref{Allvert} the vertices irreducible in the $pp$-channel ($\Gamma_{pp}$), irreducible in the $ph$-channel ($\Gamma_{ph}$) as well as the fully irreducible vertex ($\Lambda$) exhibit similar structures as the full vertex ($F$). The explanation of these features, hence, follows the discussion above. Let us just point out, that for $\Gamma_{ph}$ at $U=1$, and for both, $\Gamma_{ph}$ and $\Gamma_{pp}$ at $U=2$ sign changes appear in the vertex functions at low frequencies. This might be related to the previously observed divergences in the irreducible vertex functions.\cite{Janis01,Janis01b,VertDiv,Janis14}. For $U=2$, $\Gamma_{ph}$ even shows a square-like structure extending to a broadened cross. Within these features the vertex exhibits sign changes.
% {\bf ??? Check if this is true (not so clear from this color scale)!}. True, TR 
Considering Eq. (\ref{equ:lambda}) for $\Lambda$ the sign-changes of $\Gamma_i$ extend obviously to the fully irreducible vertex. 

As one of the two main applications of the local vertex functions of DMFT we have calculated by means of the Bethe-Salpeter equation in the $ph$ channel using the {\sl non-local} DMFT Green's functions,  the charge density wave (CDW) susceptibility $\chi(\omega,\mathbf{q})$ of the system (for the concrete calculation see Section III and Ref. \onlinecite{Georges92a}). A divergence of this susceptibility at $\omega=0$ and $\mathbf{q}=(\pi,\pi)$ signalizes the transition from a paramagnetic to a checkerboard-like charge-ordered phase (within the framework of DMFT). The values of $\chi(\omega=0,\mathbf{q}=(\pi,\pi))$ are reported in Fig. \ref{Phasediag} yielding the phase-diagram of the half-filled FKM, which agrees well with the result of the literature\cite{Antipov14}.

\FloatBarrier

\vskip 5mm
%% end tr
\section{Non-local corrections to DMFT self-energies from 1PI and DF}
\label{Sec:WIP}
\subsection{Analytic derivation of correction terms}
\label{AnDerCorTer}
%%beg gr
The dynamical mean field theory (DMFT) captures local correlations in the FKM. In this way it is possible to describe physical phenomena which are driven by local correlations such as, e.g., the Mott-Hubbard-like metal-to-insulator transition in the Falicov-Kimball model. However, DMFT cannot describe physical properties due to non-local correlations between the electrons on a finite-dimensional lattice. The latter are particularly important in the vicinity of second-order phase transitions from a paramagnetic to a spatially ordered phase. In this case the correlation length of the system can become very large and eventually diverges at the transition. While, in this situation, the order parameter itself is of course zero above the critical temperature, $T_c$, one observes strong non-local {\sl fluctuations} of this order parameter on all length-scales. 

As discussed before, the half-filled Falicov-Kimball model on a bipartite lattice exhibits an instability towards CDW ordering at low temperatures: itinerant and localized electrons arrange themselves in a checkerboard structure. Below the ordering temperature the physics of such system is to a large extent controlled by the presence of the spatial order, i.e., by the finite value of the order-parameter. This ordered phase can be described by DMFT, see Fig.\ \ref{Phasediag}.
Hallmarks of  this ordered phase can be observed however already in the paramagnetic phase slightly above the transition temperature. These effects on the other hand are not captured by the DMFT self-energy. 

Let us recall, that the effect of strong order-parameter fluctuations in this regime is very pronounced in {\sl two-particle} observables such as the charge susceptibility. The latter is strongly enhanced in the vicinity of the ordered phase and eventually diverges when approaching the transition  (see Fig. \ref{Phasediag} and Ref. \onlinecite{Antipov14}). On the other hand, the presence of the phase transition should also affect {\sl one-particle} properties of the system, in particular spectral functions and self-energies.  In order to treat such a situation from a theoretical perspective we have to include non-local correlations beyond the local ones of DMFT into the self-energy of the system. Cluster extensions like cluster dynamical mean field theory (CDMFT) \cite{clusterDMFT} or dynamical cluster approximation \cite{DCA} are able to include short range correlations within the cluster size. However, since exactly at the phase transition, the correlation length of the system diverges, a finite-cluster treatment is insufficient for a comprehensive description. In this respect, diagrammatic extensions of DMFT, which are capable of treating spatial fluctuations on all length scales, offer an alternative route for analysing the half-filled Falicov-Kimball model on the verge of charge ordering. 

Here, we have applied the dual fermion (DF) and one-particle irreducible approach (1PI) to include non-local correlations in the electronic self-energy and spectral function on top of the local ones of DMFT. The alternative, dynamical vertex approximation is considerably more difficult to implement for the FKM since it relies on the equation of motion, which in turn requires the calculation of the $c$-$f$ vertex between localized and mobile electrons. Both, DF and 1PI, start from the action of the FK model:
\begin{align}
 \label{equ:actionFK}
 \mathcal{S}_{\text{FK}}&[c^+,c,f^+,f]=\frac{1}{\beta}\sum_{\nu,\mathbf{k}}\left[-i\nu+\varepsilon_{\mathbf{k}}-\mu\right]c^+_{\mathbf{k}}(\nu)c^{\phantom{+}}_{\mathbf{k}}(\nu)\nonumber\\&
+\frac{1}{\beta}\sum_{\nu i}\left[-i\nu+\varepsilon_f\right]f^+_{{i}}(\nu)f_{i}^{\phantom{+}}(\nu) \nonumber
\\&+U\sum_i\int_0^{\beta}d\tau\;c^+_{i}(\tau)c_{i}^{\phantom{+}}(\tau)f^+_{i}(\tau)f_{i}^{\phantom{+}}(\tau),
\end{align}
where $c^{(+)}_i(\tau)$ and $f^{(+)}_i(\tau)$ represent the fermionic Grassmann fields for the itinerant and localized electrons, respectively, at lattice site $i$ and imaginary time $\tau$. $c^{(+)}_{\mathbf{k}}(\nu)$ denotes the corresponding Fourier transform of the itinerant field to frequency- and momentum-space, where $\mathbf{k}$ is a momentum vector in the first Brillouin zone and $\nu=\frac{\pi}{\beta}(2n+1),n\in\mathds{Z}$ is a Matsubara frequency at a given (inverse) temperature $\beta=1/T$.

In the spirit of DMFT we now express the actual FK model in terms of a purely local system [coined resonant level model (RLM)] for which one- and two-particle Green's functions can be obtained exactly (as was done in the previous section). To this end we replace the only non-local term in Eq. (\ref{equ:actionFK}), i.e., the lattice dispersion $\varepsilon_{\mathbf{k}}$, by a local hybridization function $\Delta(\nu)$ (e.g., the one of DMFT). It is evident that the exact action of the FKM can be then expressed in terms of this hybridization and a corrections term containing {\sl all} non-local parts of the action:
\begin{align}
 \label{equ:fkrewrite}
 \mathcal{S}_{\text{FK}}[c^+,c,f^+,f]&=\sum_i\mathcal{S}_{\text{RLM}}[c_i^+,c_i,f^+_i,f_i]-\nonumber\\&\frac{1}{\beta}\sum_{\nu,\mathbf{k}}\left[\Delta(\nu)-\varepsilon_{\mathbf{k}}\right]c^+_{\mathbf{k}}(\nu)c_{\mathbf{k}}(\nu),
\end{align}
where the action of the RLM at the lattice site $i$ can be obtained from Eq. (\ref{equ:actionFK}) by just replacing $\varepsilon_{\mathbf{k}}$ with $\Delta(\nu)$ (and, of course, omitting the sums over $i$ and $\mathbf{k}$).

\begin{figure}
	\centering
		\includegraphics[width=0.5\textwidth]{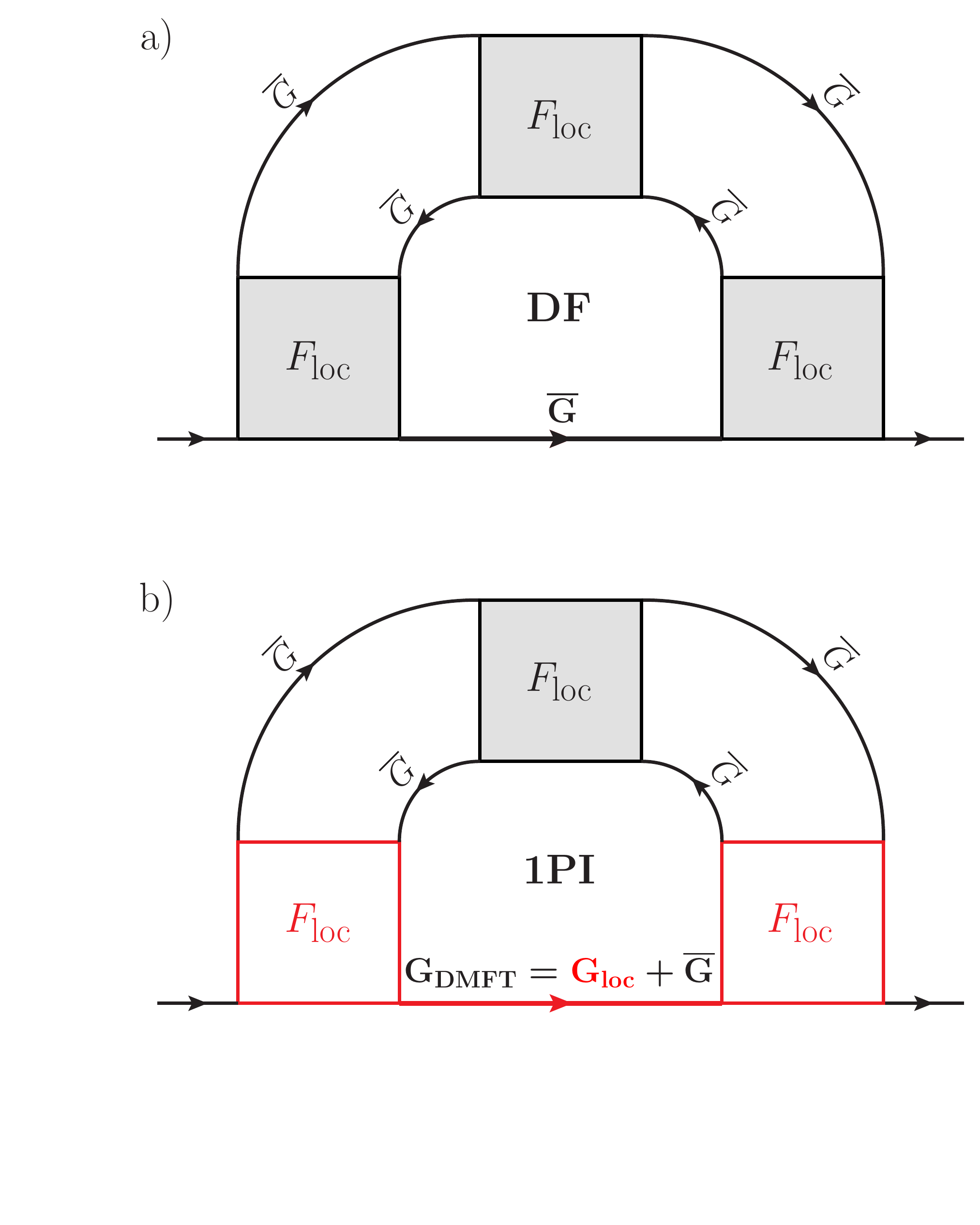}
	\caption{a) Typical third-order diagram of DF. Note, that apart from local vertex functions it contains just purely {\sl non-local} propagators $\overline{G}=G_{DMFT}-G_{loc}$. b) A diagram which can be constructed in DF only at the three-particle level (with the red part as a one-particle reducible three-particle vertex) \cite{1PI}, but is present in 1PI already at the two-particle level.}
	\label{fig:compdf1pi}
\end{figure}

The main idea of the DF and the 1PI method is now to perform a fermionic Hubbard-Stratonovich decoupling of the term in the second line of Eq. (\ref{equ:fkrewrite}):
\begin{align}
 \label{equ:cap4hsdecoupling}
 &e^{\frac{1}{\beta}\left[\Delta(\nu)-\varepsilon_{\mathbf{k}}\right]c^+_{\mathbf{k}\sigma}(\nu)c_{\mathbf{k}\sigma}(\nu)}\propto\int d\widetilde{c}^+_{\mathbf{k}\sigma}(\nu)d\widetilde{c}_{\mathbf{k}\sigma}(\nu)\nonumber\\&e^{\pm\frac{1}{\sqrt{\beta}}\left[\Delta(\nu)-\varepsilon_{\mathbf{k}}\right]^{\frac{1}{2}}B_{\mathbf{k}\sigma}(\nu)\left[c^+_{\mathbf{k}\sigma}(\nu)\widetilde{c}_{\mathbf{k}\sigma}(\nu)+\widetilde{c}^+_{\mathbf{k}\sigma}(\nu)c_{\mathbf{k}\sigma}(\nu)\right]}\nonumber\\[-0.2cm]&\times e^{-\left[B_{\mathbf{k}\sigma}(\nu)\right]^2\widetilde{c}^+_{\mathbf{k}\sigma}(\nu)\widetilde{c}_{\mathbf{k}\sigma}(\nu)},
\end{align}
where the $\widetilde{c}^{(+)}$ are the Hubbard-Stratonovich fields, which are coined ``dual fermions'' in the framework of the DF theory. Choosing $B_{\mathbf{k}\sigma}(\nu)=\left[G_{\text{loc}}(\nu)\right]^{-1}\left[\Delta(\nu)-\varepsilon_{\mathbf{k}}\right]^{-\frac{1}{2}}$ allows us to rewrite the term in Eq. (\ref{equ:cap4hsdecoupling}) which couples the real and the dual fermions from a sum over momentum to a sum over real space:
\begin{equation}
\label{equ:cap4klatticesum}
\sum_{\nu,\mathbf{k}}c^+_{\mathbf{k}}(\nu)\widetilde{c}_{\mathbf{k}}^{\phantom{+}}(\nu)+\widetilde{c}^+_{\mathbf{k}}(\nu)c_{\mathbf{k}}(\nu)=\sum_{\nu,i}c^+_{i}(\nu)\widetilde{c}_{i}^{\phantom{+}}(\nu)+\widetilde{c}^+_{i}(\nu)c_{i}^{\phantom{+}}(\nu)
\end{equation}
After these transformations the action regarding only the original (physical) fields becomes diagonal in real space and, hence, the original fermions can be integrated out locally (i.e., separately for each lattice site). In this way, one obtains an effective action for the new fermions, whose free propagator is just given by the difference of the DMFT Green's function and its local counterpart, while the interaction between these new particles are just the one-, two- and more-particle local (connected) vertex functions of DMFT. Hence, the new theory contains already in its lowest order diagrammatic expansion all local correlations of DMFT via the corresponding DMFT self-energy, while non-local corrections can by constructed diagrammatically by means of the above mentioned propagator and the local DMFT vertex functions. A typical (third order) diagram of DF is shown in Fig. \ref{fig:compdf1pi}. 

It is important to note that in the DF theory the {\sl full} two- and more-particle vertex functions act as interaction between the dual electrons. Apart from one-particle irreducible (1PI) contributions these vertices contain also {\sl one-particle reducible} parts. This leads to two main difficulties in the DF approach: (i) The diagrammatic sums performed within DF will in general contain {\sl one-particle reducible} contributions to the (dual) self-energy which have to be removed by corresponding counterterms\cite{katanin}; (ii) In almost all diagrammatic extensions of DMFT, only local {\sl two-particle} vertex functions are taken into account for constructing non-local corrections to the DMFT self-energy. Three- and more-particle local vertices are usually neglected. Within such an approximation DF {\sl does not} generate all diagrams which could be constructed from these local two-particle vertex functions. This is illustrated exemplary in Fig. \ref{fig:compdf1pi}: The diagram shown in b) is {\sl not} contained in the DF theory when restricted to two-particle vertices. In fact, it contains a purely local propagator which is not available in the dual fermion theory. Specifically, the part marked in red represents a one-particle reducible contribution to the local three-particle vertex. Hence, the diagram in the Fig. \ref{fig:compdf1pi} b) can be constructed in DF only when including local {\sl three particle} vertex-functions. One the other hand, this diagram consists -apart from local and non-local DMFT Green's functions- only of two-particle vertex functions. Hence, it would be desirable to include it already at the two-particle level in the theory. 

The problem mentioned above can be avoided when excluding one-particle reducible diagrams from the theory. This can be done in the standard way by performing a Legendre transform on the generating functional\cite{1PI}. As a result, the interaction between the new fields is given only by the one-particle irreducible vertex functions. On the other hand, one can show that for these new fields a purely local propagator is available. Hence, the diagram in Fig. \ref{fig:compdf1pi} is generated in this new 1PI-approach already at the two-particle level. From the considerations above it is evident that the set of diagrams included in the DF approach represents a subset of the diagrams taken into account in the 1PI theory when restricting both methods to the two-particle local vertex functions.

%{\bf However, for the peculiar model studied in this work, i.e., the half-filled FKM, the one-particle reducible three-particle vertex vanishes.\cite{Antipov14} This is not the case for the 1PI three-particle vertex. There is  a cancellation of the  counterterms with the 1PI diagrams for the three-particle vertex. Hence, the first correction terms to the  two-particle vertex DF theory vanish, suggesting that for the special case of the half-filled FKM the DF results might be more accurate.}

%{\bf Ich würde hier folgenden Text vorschlagen:
However, for the peculiar model studied in this work, i.e., the half-filled FKM, the full three-particle vertex vanishes\cite{Antipov14} due to a perfect cancellation between one-particle reducible and 1PI contributions. As in 1PI only the 1PI vertex is considered in the theory, such a cancellation does not take place in the effective interaction of the 1PI approach. This suggests, that for this special case DF results might be more accurate.
%}

For the formal derivation of the DF and the 1PI expressions for the non-local corrections to the self-energy we refer the reader to the literature \cite{DualFermion,1PI}. Here, we present just the final equations for the self-energy corrections which are obtained from ladder contributions:
\begin{subequations}
 \label{equ:cap4sigma1sigma2}
\begin{align}
 &\Sigma_{1,k}=-\frac{2}{\beta^2}\sum_{k'q}F_{\text{loc}}^{\nu\nu'\omega}\overline{G}_{k'}\overline{G}_{k'+q}F_{\mathbf{q}}^{\nu'\nu\omega}\overline{G}_{k+q}-\Sigma_{1,k}^{(2)},\label{equ:cap4sigam1}\\
 &\Sigma_{2,k}=-\frac{2}{\beta^2}\sum_{k'q}F_{\text{loc}}^{\nu\nu'\omega}\overline{G}_{k'}\overline{G}_{k'+q}F_{\mathbf{q}}^{\nu'\nu\omega}G_{\text{loc},\nu+\omega},\label{equ:cap4sigam2}.
\end{align}
\end{subequations}
Note that
$\Sigma_{2,k}\equiv\Sigma_{2,\nu}$ is, in fact, $k$-independent (1PI correction to the {\sl local} self-energy due to non-local corrections). $\Sigma_{1,k}^{(2)}$ is defined as:
\begin{equation}
 \label{equ:cap4defsigma2nd}
 \Sigma_{1,k}^{(2)}=-\frac{1}{\beta^2}\sum_{k'q}F_{\text{loc}}^{\nu\nu'\omega}\overline{G}_{k'}\overline{G}_{k'+q}F_{\text{loc}}^{\nu'\nu\omega}\overline{G}_{k+q}, 
\end{equation}
The latter contribution accounts for the double-counting of the second-order (in $F_{\text{loc}}$) diagram due to the indistinguishability of identical particles. Finally, the vertex $F_{\mathbf{q}}^{\nu\nu'\omega}$ is constructed from ladder diagrams consisting of local two-particle vertices and non-local Green's functions of DMFT:
\begin{equation}
 \label{equ:cap4deffq}
 F_{\mathbf{q}}^{\nu\nu'\omega}=F_{\text{loc}}^{\nu\nu'\omega}+\frac{1}{\beta}\sum_{k_1}F_{\text{loc}}^{\nu\nu_1\omega}\overline{G}_{k_1}\overline{G}_{k_1+q}F_{S\mathbf{q}}^{\nu_1\nu'\omega},
\end{equation}
where $F_{\text{loc},r}^{\nu\nu'\omega}$ defines here the local two-particle vertex function of DMFT. Let us stress, that in order to capture the above discussed physics of long-range correlations it is absolutely necessary to consider ladder-like diagrams which are capable of describing these fluctuations on all length scales. 

A peculiarity of the FKM is that, similarly as for the local vertex functions in the previous section, the correction formulas for the DMFT self-energy can be given (semi)-analytically. Indeed, after some algebraic transformations, analogously to those in section II A we obtain:
%end gr
%start tr
\begin{multline}
\Sigma^{(1)}_1 (\nu , k ) = 2 \sum_q C_q \dfrac{a^2 (\nu)}{ \left( 1 - a^2 (\nu) \Chired^{\nu , \nu }(q) \right)^2 } \Gred ( \nu , k + q ) -  \\ 2 \sum_{q , \nu_1} \dfrac{a (\nu) a (\nu_1) }{ 1 - a (\nu) a (\nu_1) \Chired^{\nu , \nu_1}(q) } \Gred (\nu_1 , k + q ) 
\label{215}
\end{multline}
\begin{multline}
\Sigma^{(2)}_1 (\nu , k ) = 2 \sum_{\nu_1 , q} a^2 ( \nu ) a^2 ( \nu_1 ) \Chired^{\nu_1 , \nu_1} (q) \Gred ( \nu , k + q ) -  \\ 2 \sum_{ q } a^4 (\nu ) \Chired^{\nu , \nu} (q) \Gred ( \nu , k + q ) 
\label{216}
\end{multline}
\begin{multline}
\Sigma_2 (\nu ) = 2 \sum_q C_q \dfrac{a^2 (\nu)}{ \left( 1 - a^2 (\nu) \dfrac{1}{\beta} \Chired^{\nu , \nu}(q) \right)^2 } G_{ loc } ( \nu ) -  \\ 2 \sum_{q , \nu_1} \dfrac{a (\nu) a (\nu_1) }{ 1 - a (\nu) a (\nu_1) \Chired^{\nu , \nu}(q) } G_{loc} (\nu_1 ) + \\
\dfrac{2}{\beta} \sum_{ \nu_1 } a( \nu_1 ) G_{loc} ( \nu_1 ) a ( \nu ) - \dfrac{2}{ \beta } a^2 ( \nu ) G_{ loc } ( \nu ) .
\label{217}
\end{multline}

Here $\Chired^{\nu , \nu'}(q)$ is defined as 
\begin{equation}
\Chired^{\nu , \nu'}(q) = \sum_k \Gred(\nu , k) \Gred (\nu' , k + q)
\end{equation}
and $C_q$ is given by
\begin{equation}
C_q = \left( 1 + \sum_\nu \dfrac{a^2(\nu) \Chired^{\nu , \nu}(q)}{1 - a^2(\nu) \Chired^{\nu , \nu}(q)} \right)^{-1} .
\end{equation}
$\Sigma^{(1)}_1$, $\Sigma^{(2)}_1$ and $\Sigma_2$ represents the single contributions of the 1PI theory to the corrections of the DMFT self-energy originating from non-local correlations in the system. $\Sigma^{(1)}_1$ and $\Sigma^{(2)}_1$ represents exactly the DF corrections corresponding to diagrams of the type shown in Fig. \ref{fig:compdf1pi} a). In the 1PI approach on the other hand, also a purely local propagator is available giving rise to the correction $\Sigma_2$ of the DMFT self-energy. A typical diagram contributing to $\Sigma_2$ is depicted in Fig. \ref{fig:compdf1pi} b). 

Let us point out that the sum of $\Sigma_1^{(1)}$ and $\Sigma_1^{(2)}$ does not represent the final self-energy correction to the DMFT self-energy obtained from dual fermion: In fact they represent just the corrections for the dual particles and, hence, have to be transformed back to the space of real electrons\cite{df}. Finally, we want to mention, that the results presented in the following are obtained by so-called ``one-shot'' calculations, i.e., no self-consistency has been performed in the DF theory\cite{DualFermion} and no $\lambda$-corrections have been applied to the 1PI-results\cite{1PI}.

%Let us turn our attention to the denominator appearing in $C_q$ and $\Sigma^{(1)}_1$, which can eventually become very small and, hence, may lead to large corrections to the self-energy as we will see in the next section. Let us finally note that the DF correction to the DMFT self-energy is not given directly by $\Sigma^{(1)}_1+\Sigma^{(2)}_1$ but still has to be transformed from the dual space back to the space of real electrons\cite{DualFermion}.

\subsection{Numerical results in two dimensions}
\label{NumRes2D}
In this section we present numerical results obtained for the self-energies of the DF and the 1PI approach for $U=1.0$ and $U=2.0$ and for two temperatures respectively. To this end we have evaluated Eqs. (\ref{215})-(\ref{217}) numerically using $120$ fermionic frequencies for the Matsubara summations and $160 \, \mathbf{k}$-points (in each direction) for performing the momentum-integrals. All calculations presented in this section have been performed at half-filling.

%{\bf I am not sure if we need the following paragraph}
%It is well known that the FKM does not describe a Fermi liquid and, hence, its DMFT self-energy (on the real frequency axis) is not Fermi liquid like\cite{Freericks03}. Specifically, the the slope of its real part is positive at $\omega=0$ (instead of negative as for a Fermi liquid) leading to a value of the quasiparticle renormalization factor $Z>1$. Hence, in the renormalized dispersion $\widetilde{\varepsilon}_{\mathbf{k}}=Z\varepsilon_{\mathbf{k}}$ states are shifted away from the Fermi level $\varepsilon_{k_F}=0$. In addition to this shift of energy levels away from the Fermi level driven by the real part of the self-energy, the imaginary part of $\Sigma(\nu)$ leads to to a strong decay of the remaining quasiparticles at the Fermi level. In the spectral function of the system such a behavior of the self-energy translates into a gap at the Fermi level rendering the system less metallic. With increasing values of the interaction $U$ this effect gets stronger until the system reaches the so-called Mott-insulating phase where the system is fully gapped (at $T=0$). Our numerical calculations, however, are of course performed for Matsubara rather than real frequencies. There the increase of the local correlations with $U$ in the DMFT self-energy is represented by an increase of the Matsubara self-energy at the lowest Matsubara frequencies. 

\begin{figure}
\includegraphics[width=.95\linewidth]{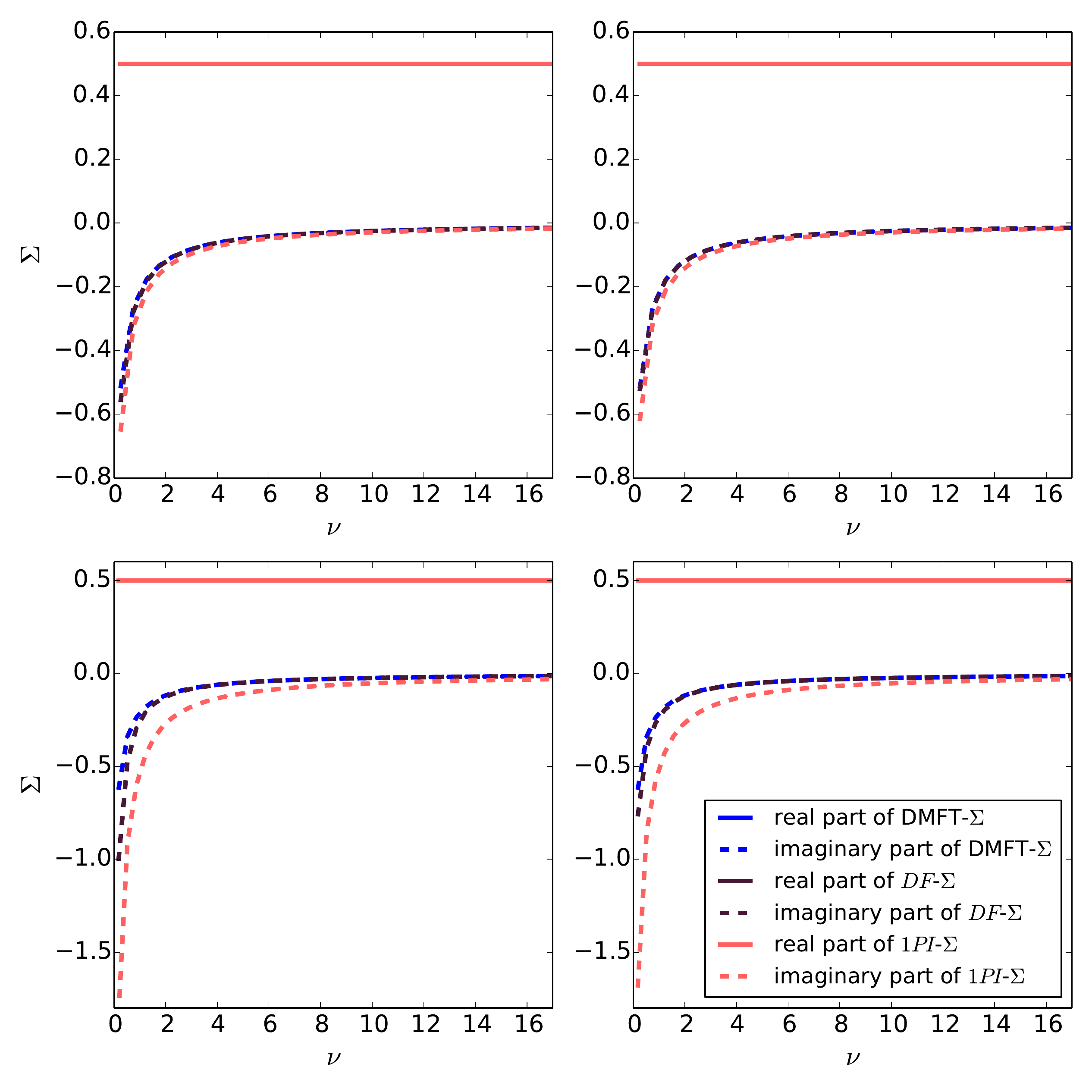}
\caption{\label{Komp}DF and 1PI self energies for the $k$-points $(\pi,0)$ (left) and  $(\pi / 2,\pi / 2)$ (right) at $U=1$ for $T = 0.055$ (bottom) and $T = 0.08$ (top) }
\end{figure}

In Fig. \ref{Komp} we compare the DMFT, DF and 1PI self-energies for $U=1$ at two different temperatures, $T=0.08$ and $T=0.055$. One can clearly see that the non-local correlations captured by DF and 1PI strongly enhance the imaginary part of the DMFT self-energy (note that  at half filling the real part on the Matsubara axis is just given by the Hartree term). As expected, at a higher temperature $T=0.08$ this effect is relatively moderate and also differences between the nodal [$\mathbf{k}=(\pi/2,\pi/2)$] and the antinodal [$\mathbf{k}=(\pi,0)$] $\mathbf k$-point on the Fermi surface are insignificant. This is consistent with the fact, that at high temperatures the system exhibits a mean-field-like behaviour with only small corrections to DMFT. Consistent with this, also the differences in the self-energies between 1PI and DF are relatively small.

At a lower temperature $T=0.055$ the corrections of 1PI and DF to DMFT become much larger as it can be seen in the second row of Fig. \ref{Komp}. That means that non-local correlations strongly affect the self-energy at this set of parameters. According to the phase diagram in Fig. \ref{Phasediag} the considered data point is already (quite) close to the CDW phase transition; the correlation length extracted from the non-local ph-ladder is given by $\xi=2.63$. Already for such a $\xi$, the corresponding charge susceptibilities is large. Since the correlation length and susceptibilities diverge, non-local correlations will be even much larger close to  $T_c$. The susceptibility  contributes to the DF and 1PI self-energies via the terms in the first row of Eqs. (\ref{215}) and (\ref{217}). Hence both, DF and 1PI, demonstrate that strong charge fluctuations enhance the self-energy in the vicinity of the CDW phase transition of DMFT in the two-dimensional FKM. Our results are also consistent with the observation that  non-local corrections are substantially stronger at the anti-nodal than at the nodal point: At the anti-nodal point CDW fluctuations even stronger enhance the self-energy due to the presence of a van Hove singularity in the density of state of the bipartite square lattice. 

As for difference between DF and 1PI one can see in Eqs. (\ref{215})-(\ref{217}) that in 1PI in addition to the purely non-local propagator $\overline{G}(\nu,\mathbf{k})$, which is rather small, also a purely local DMFT propagator $G_{\text{loc}}(\nu)$ appears in the equations, cf.\ Fig. \ref{fig:compdf1pi} b). 
The latter can be seen as part of a reducible local three-particle vertex of DMFT, which appear in DF only when considering three-particle vertex functions explicitly. The difference  will be reduced if a $\lambda$-correction\cite{Katanin09} was included in 1PI as this shifts the CDW divergence towards lower temperatures. More details about the relation between 1PI and DF are discussed in Ref. \onlinecite{1PI}. 
\begin{figure}
\includegraphics[width=.95\linewidth]{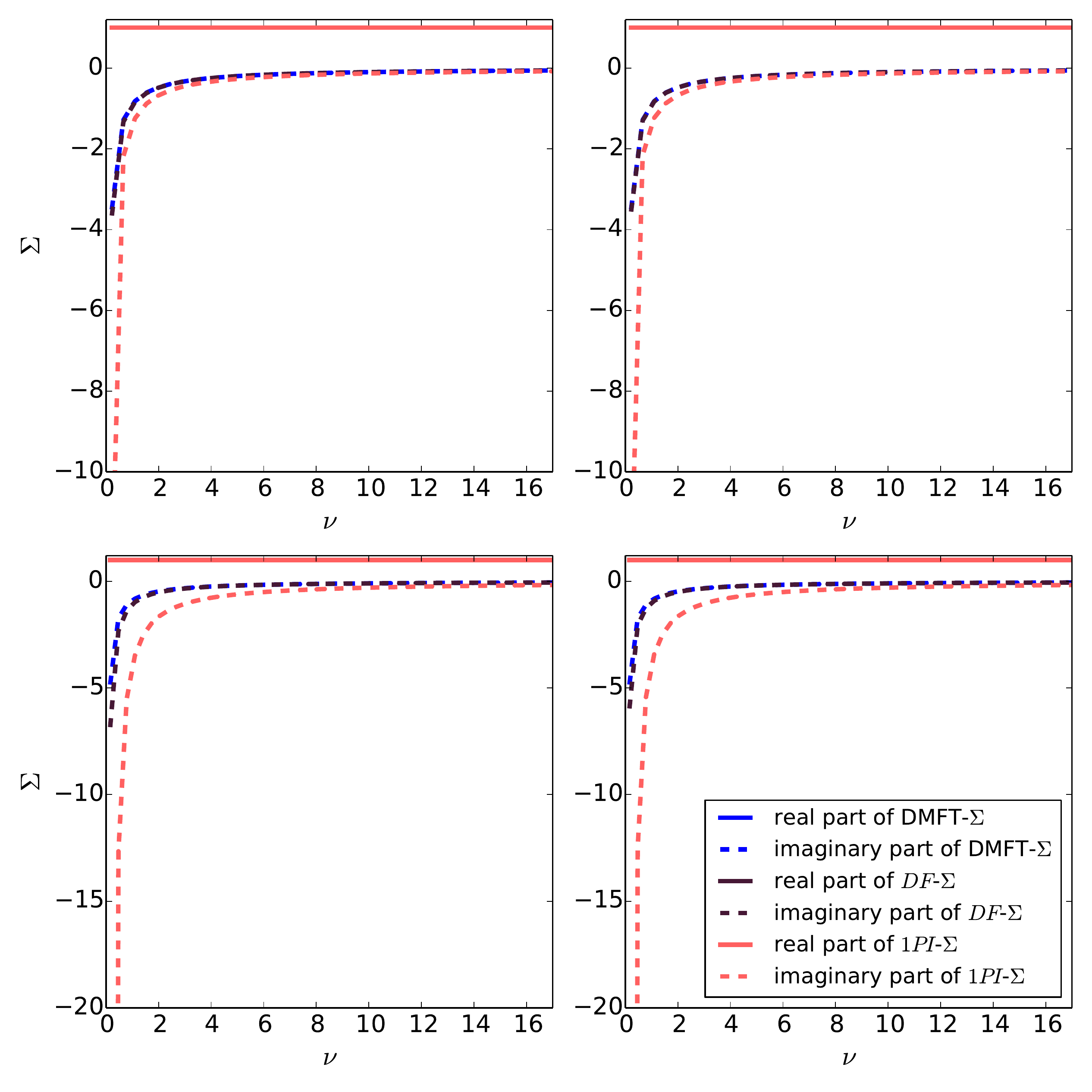}
\caption{\label{Gomp}DF and 1PI self energies for the $k$-points $(\pi,0)$ (left) and  $(\pi / 2,\pi / 2)$ (right) at $U=2$ for $T = 0.05$ (bottom) and $T = 0.07$ (top) }
\end{figure}

The situation described above is even more pronounced for the larger coupling,  $U=2$, as illustrated in Fig. \ref{Gomp}. One observes large corrections to the DMFT self-energy from 1PI and DF whereas again the effect of non-local correlations is larger at lower temperatures (close to the charge-ordering phase transition) and for the anti-nodal point on the Fermi surface. As for the difference between DF and 1PI one can see (similar as for $U=1$) that the corrections of the 1PI method are much larger than that of the DF ones.

In order to get a better physical insight into the meaning of our results we performed an analytic continuation of our self-energy data to the real axis by means of a Pad\'e fit to frequencies slightly above the real axis (${\rm Im} \, \omega = 0.1$) 
%???korrekt???in the caption it was + should not be???. checked, TR 
From the corresponding self-energy on the real axis we then obtain the spectral functions for the system from DMFT, DF and 1PI. In Fig. \ref{K2Gs} 
%{\bf Add DMFT Spectra! ???done???} Done, TR
we compare the $\mathbf{k}$-resolved spectral functions of DMFT, DF and 1PI for $U=1$ and $T=0.055$ for the nodal (right) and the antinodal (left) point, respectively. For all methods we observe a gap at the Fermi level whereas its depth is larger in the 1PI data than in DMFT. This is of course expected from the enhanced behaviour of the 1PI (and DF) self-energy on the Matsubara axis compared to DMFT. In the DMFT spectrum we clearly observe two peaks corresponding to the formation of two Hubbard-like bands, consistent with the DMFT metal-insulator transition at $U_c=1$. 
%??correct??. Now it is, TR

\begin{figure}
\includegraphics[width=.9\linewidth]{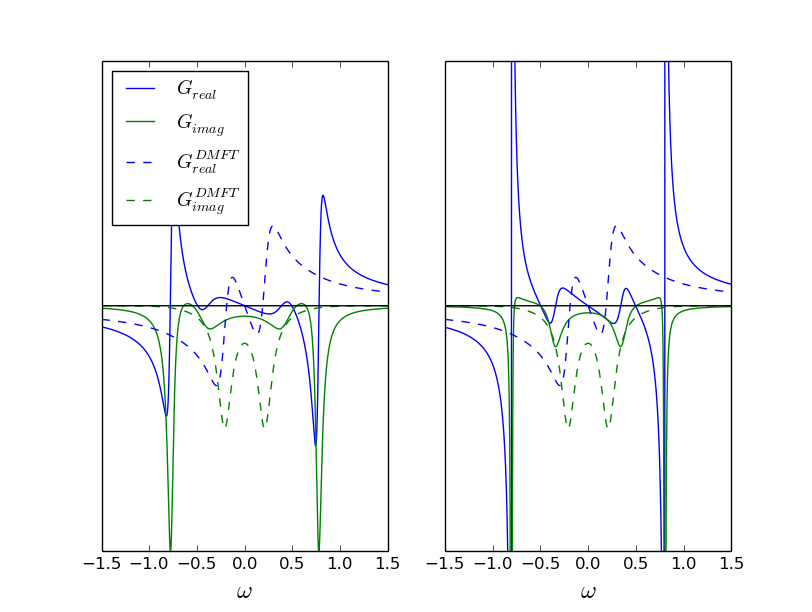}
\caption{\label{K1Gs}Green's functions for the $k$-points $(\pi,0)$ (left) and $(\pi / 2,\pi / 2)$ (right) at $U=1$ for $T = 0.055$. We attribute small non-analyticities (positive imaginary parts) to the Pad\'e-fit.}
\end{figure}

\begin{figure}
\includegraphics[width=.9\linewidth]{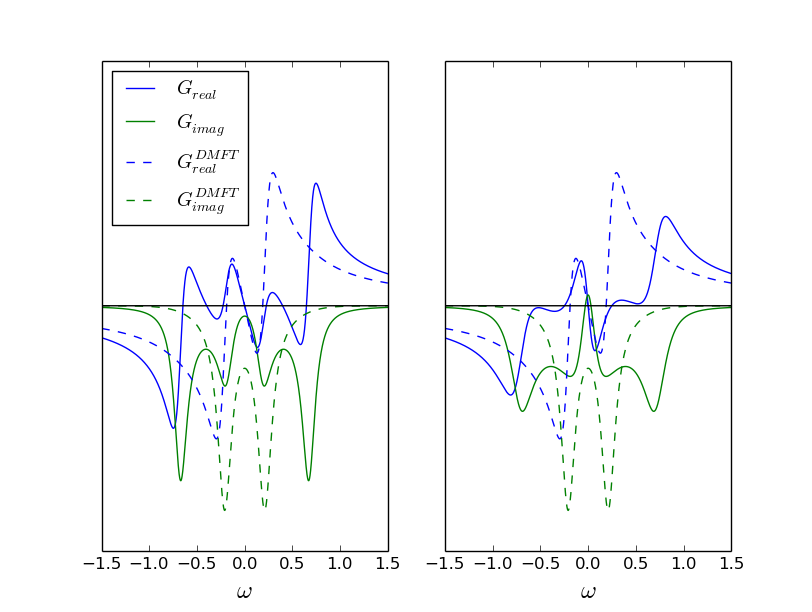}
\caption{\label{K2Gs}Green's functions for the $k$-points $(\pi,0)$ (left) and $(\pi / 2,\pi / 2)$ (right) at $U=1$ for $T = 0.07$. We attribute small non-analyticities (positive imaginary parts) to the Pad\'e-fit.}
\end{figure}

Interestingly, in the 1PI results we find a four-peak structure. While 
it is difficult to  exclude irrevocably  that 
this feature is not an artifact of the analytic continuation, we find that the four-peak  structure is rather stable and present also for different parameter sets close to the CDW transition of DMFT. A possible interpretation of such features might be that each of the two DMFT-split bands, originating purely from local correlations, is further split into two peaks by precursors of the CDW 
ordering, i.e., originating from  non-local correlations. 
%Such a splitting of the density of states due to charge-ordering fluctuations can be already observed in a static mean-field treatment of the FKM Hamiltonian in the symmetry broken phase. Hence, the four-peak structure of the spectral function might possibly be interpreted as a precursor of the ordered phase. 
This assertion is supported by the spectral functions obtained at $U=2$ and $T=0.055$ in Fig. \ref{K1Gs}: Here the four-peak structure of the 1PI spectral function is even more pronounced, while in DMFT only two peaks representing the two separated bands can be observed.

A  similar  four-peak feature in the spectral function is visible  in Monte-Carlo data for the FKM (Figure 21 (f) of Ref.\ \onlinecite{FKMC}, Refs.\ \onlinecite{PokorAddcite1} and \onlinecite{PokornUnpub} ) and dynamical cluster approximation (DCA) \cite{PokorAddcite2}. A four-peak structure 
 has also been reported
 in (semi)analytical calculations for the Hubbard model\cite{Tremblay}, which
attributed this structure to a mixture of an antiferromagnetic and a Hubbard band splitting. 

Turning to our analytical expression,  we associate the four peak structure to a combination of the DMFT (and strong coupling) pole at $\nu=0$ leading to the DMFT band splitting, and additional poles at a finite $\pm \nu$. The latter pole develop if the denominator  in Eq.\ \ref{217}, $1-a (\nu) a (\nu_1) \Chired^{\nu , \nu}(q)$, approaches zero. Since this denominator originates from the geometric series of the  charge susceptibility  particle-hole ladder diagrams, we can identify it with non-local CDW fluctuations.

\begin{figure}[tb]
\includegraphics[width=.9\linewidth]{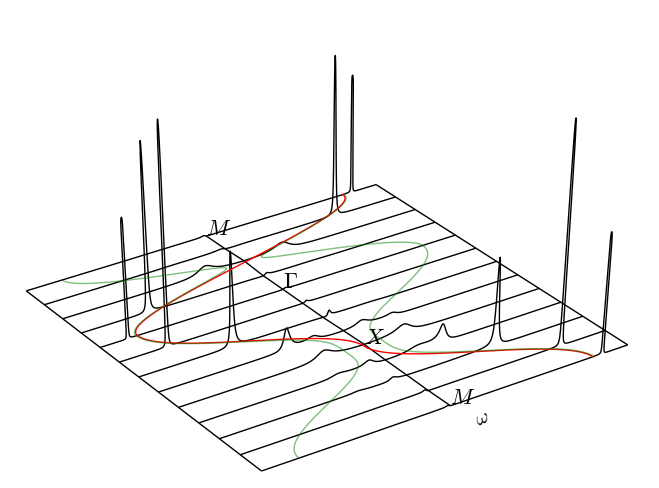}
\caption{\label{Subspectra}Spectral functions along a path through the Brillouin zone as a function of the real frequency $\omega$ for $U=1$ and $T = 0.07$. For comparison, the  eigenenergies of the square lattice (red/dark) and a simple non-interacting system with on-site energies $-U/2$ and $U/2$ on a checkerboard (green/light) are plotted.}
\end{figure}

Finally, in Fig. \ref{Subspectra} we show spectral functions for $U=1$ at $\mathbf{k}$-points on and away from the Fermi level. One can see that the spectral weight away from the Fermi energy (M and $\Gamma$ points) exhibits a strong peak at precisely $\varepsilon_k$ (red line) indicating that the the imaginary part of the self-energy is rather small there. On the other hand, at the anti-nodal (X) point we observe a strong suppression due to the large value of the self-energy and a splitting into a four-peak structure. The splitting at the X-point 
reminds of the typical  CDW (or antiferromagnetic) splitting  or the precursor thereof, cf.\ green line.

%??? did not understand phase space argument which usually is alrger scattering away from E_F ??? took it away ??? This behavior can be easily understood by a phase-space argument: The suppression of spectral weight in 1PI occurs due to the scattering of the electrons at non-local charge fluctuations. However, such scattering events can only take place if enough space for the scattered particle is available. This is indeed the case for an electron at the Fermi surface but it is not true for electrons far away from the Fermi surface.

\section{Conclusion}
\label{Conclusio}
The derivation of closed-form expressions for all vertices of the Falicov-Kimball model allows us to calculate non-local self-energy corrections to DMFT analytically, within the  DF and 1PI approach. 

%This provides us with a straightforward way of identifying the algebraic origin of singularities in the irreducible vertex and other features arising from  non-local correlations. 

 In a pioneering work, Ref.\ \onlinecite{Antipov14}, it was already shown that the dual fermion critical exponents for the FKM are of the  Ising universality class. Beyond  Ref.\ \onlinecite{Antipov14}, we show how charge fluctuations effect the paramagnetic spectral functions. These non-local correlations lead to a more insulating solution with a, compared to DMFT, larger splitting of the lower and upper band. Our results also indicate a four-peak structure of the ${\mathbf k}$-resolved spectral function in parts of the Brillouin zone. 
As a physical explanation we propose a dynamical mixture  of the DMFT metal-insulator transition 
caused by local correlations, and 
non-local  checkerboard CDW correlations.
Similar  four-peak features in the spectral function can be identified in Monte-Carlo studies for the FKM \cite{FKMC} and have been observed  in (semi)analytical calculations for the Hubbard model\cite{Tremblay}.

{\sl Acknowledgements.} We thank Prabuddha Sanyal, Veljko Zlati\'c, Andrey Antipov and Alessandro Toschi for very helpful discussions. Some of the plots were made using the matplotlib \cite{Matplotlib} plotting library for python. Financial support is acknowledged from the Austrian Science Fund (FWF) through I-610-N16 as part of the DFG research unit FOR 1346 (GR,KH) and the  European Research Council under the European Union's Seventh Framework Program (FP/2007-2013)/ERC through grant agreement n.\ 306447 (TR, KH).

\FloatBarrier

\end{document}